\documentclass[preprint,12pt]{elsarticle}

\usepackage{amssymb}
\usepackage{gensymb}
\usepackage{hyperref}
\usepackage{subcaption}
\hypersetup{colorlinks,urlcolor=blue}

\journal{Colloids and Surfaces A}

\begin{document}

\begin{frontmatter}



\title{How bees and foams respond to curved confinement: level set boundary representations in the Surface Evolver}


\author{A. Mughal}

\address{Institute of Mathematics, Physics and Computer Science, Aberystwyth University, Penglais, Aberystwyth, Ceredigion, Wales, SY23 3BZ, United Kingdom}

\address{Theoretische Physik, Friedrich-Alexander-Universit\"{a}t Erlangen-N\"{u}rnberg - Staudtstr. 7, 91058 Erlangen, Germany}

\author{T. Libertiny}
\address{Studio Libertiny, Schiestraat 46, 3013 AH Rotterdam, The Netherlands}

\author{G. E. Schr\"{o}der-Turk}

\address{Murdoch University, School of Engineering and Information Technology, Mathematics and Statistics, Murdoch, WA6162, Australia}

\address{Theoretische Physik, Friedrich-Alexander-Universit\"{a}t Erlangen-N\"{u}rnberg - Staudtstr. 7, 91058 Erlangen, Germany}

\begin{abstract}
We present a Surface Evolver framework for simulating single bubbles and multicellular foams trapped between curved parallel surfaces. We are able to explore a range of geometries using level set constraints to model the bounding surfaces. Unlike previous work, in which the bounding surfaces are flat (the so called Hele-Shaw geometry), we consider surfaces with non-vanishing Gaussian curvature, specifically the sphere, the torus and the Schwarz Primitive-surface. In the case of multi-cellular foams - our method is to first distribute a set of $N$ points evenly over the surface (using an energy minimisation approach), these seed points are then used to generate a Voronoi partition, that is clipped to the confining space, which in turn forms the basis of a Surface Evolver simulation. In addition we describe our experimental attempt to generate a honeycomb on a negatively curved surface, by trapping bees between two Schwarz Primitive-surfaces. Our aim is to understand how bees adapt the usual hexagonal motif of the honeycomb to cope with a curved surface. To our knowledge this is the first time that an attempt has been made to realise a biological cellular structure of this type.

\end{abstract}


\end{frontmatter}

% \linenumbers

\section{Introduction}
\label{}

\begin{figure} 
\begin{center}
\centering
\includegraphics[width=1.0\columnwidth ]{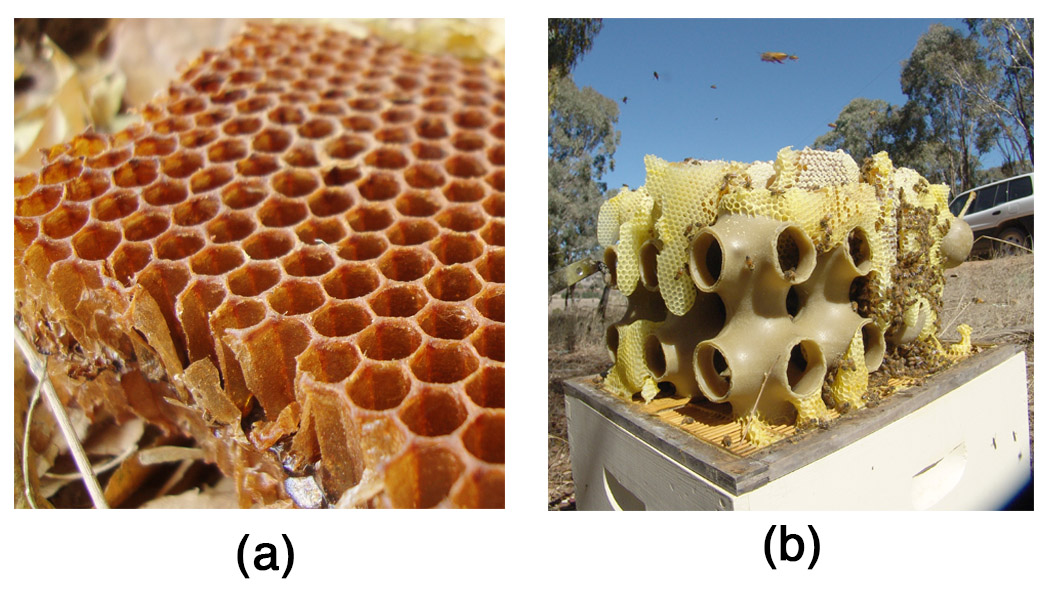}
\caption{(a) Part of a honeycomb showing some of the three-dimensional structure of the cells. Source: \href{https://commons.wikimedia.org/wiki/File:Honeycomb_15_03_2012.jpg}{Wikimedia Commons} (b) An initial attempt to build a honeycomb on a curved surface - by placing a Scharwz-P surface inside the hive. It can be seen that the bees do not confine the comb to the template surface and instead build around it. Image credit: Dr Tim Wetherell}
\label{comb}
\end{center}
\end{figure}

Since antiquity the regular hexagonal arrangement of bee honeycombs (see Fig \ref{comb}a) has been an object of fascination. Indeed it was the Roman scholar Marcus Terentius Varro who proposed in 36BC \cite{varro1889rerum} what is now known as the Honeycomb Conjecture: that hexagons are best way to divide a surface into regions of equal area with the least total perimeter.  The conjecture was eventually proven in 1999 by Thomas Hales \cite{hales2001honeycomb}. However, the debate as the exact reasons as to why bees build such striking structures continues. While the need to minimise the total amount of wax used is part of it (according to some estimates eight ounces of honey are consumed to produce one ounce of wax \cite{coggshall1984beeswax}) the reality is more complicated than the simple model considered in the Honeycomb Conjecture.

One complication is that a honeycomb is not a single sided object. Two layers of cells meet on a surface to make a single comb. The layers are slightly off-set from each other giving rise to a faceted wall between cells from opposing sides of the comb \cite{weaire2008pursuit}. Surprisingly this arrangement (which is realised in wild honeycombs) is not the optimal least-wax-using solution \cite{toth1964bees}. Further deviations from the ideal arrangement also arise from the fact that cells are not always perfectly hexagonal; distorted cells are often observed - especially when the bees encounter an obstacle in building the comb.

The exact process by which the honeycomb is built also remains largely a mystery. There are two opposing schools of thought. The first supposes that the bees actively mold the arrangement of the cells until a regular pattern is achieved \cite{bauer2013hexagonal}; in his account of the process Darwin proposed that the bees build the honeycomb by forming rough walls and refining them. The second school, dating back to at least D'Arcy Wentworth Thompson, rejects this iterative model \cite{weaire2008pursuit}. They suggests that simple physical forces play a dominating role: the bees begin by building cells as a close packed (triangular) arrangement of cylindrical precursors, but the heat inside the hive keeps the wax molten allowing it to flow and minimise surface tension. Thus mechanical tension between adjacent cells leads to a regular structure in a self-organised manner \cite{karihaloo2013honeybee, pirk2004honeybee}. Attempts at real time imagining of the process in action have been made \cite{bauer2013hexagonal} but the debate remain unresolved. 

In light of this, we propose a novel experiment to probe the response of the bees when they are subjected to a constraint. We force the hive to build a honeycomb under conditions where the usual hexagonal arrangement is known not to be the optimal solution. Here we describe our experimental attempt to build a honeycomb within the confines of two parallel negatively curved Schwarz Primitive surfaces (see Fig \ref{comb}b for our previous attempt involving only a single Schwarz-P surface). Such surfaces have a number of interesting properties including the fact that they have zero mean curvature everywhere and are periodic in all three directions  \cite{hyde2008short, grosse1997gyroid}. However, our interest in such surfaces is motivated by the fact that they present a unique challenge for any hive trying to build on it: the Gaussian curvature is continuously changing and as such any patch of honeycomb built by the bees would be frustrated if they try to extend the local arrangement out over the rest of the surface.

Although our attempt to build such a negatively curved honeycomb is only partially successful - nevertheless it may add something to the continuing debate. The question we seek to answer is: are the bees hard wired through evolution to build hexagons or do they optimise the arrangement ``on the fly'' as they are building? It is well known that a curved surface cannot be tiled entirely by hexagons alone \cite{bowick2009two}. Thus if the bees are forced to build a honeycomb within the confines of an unusual (although completely well understood) geometry what will be the result? Will the hive adapt the regular honeycomb structure by including topological defects (i.e. cells that have more or less than six sides). Such defects are often observed in the self-assembly of two-dimensional crystals on curved surfaces \cite{bowick2009two}, and their presence may hint at the underlying building processes at work.

This leads to a second problem which we also discuss here: what is the least perimeter way to enclose a number of equal volume cells between parallel curved surfaces? To answer such questions we turn to a related system that may provide further insight: monodisperse foams in confined geometries. Due to surface tension the energy of a soap foam depends directly on its on the interfacial area separating bubbles.  As such dry foams  are a useful means to investigate  area minimising partitions. Indeed the similarity between the hexagonal bee honeycomb and geometry of the hexagonal two-dimensional dry foam has long been recognised. Thus it is natural to ask how the foam structure is modified in the presence of curvature to compare with bee honeycombs between curved surfaces.

In recent years there has been a considerable progress in the study of two dimensional (2D) foams obtained by trapping bubbles between two parallel flat glass plates (the so called Hele-Shaw cell) \cite{cox2006shear, cox2008structure, drenckhan2010monodisperse, stevenson2012foam}. Such foams are not strictly two-dimensional and are more correctly referred to as being quasi-2D. However, provided the distance between the bounding plates is less than the bubble size then the effect of the third dimension can be neglected \cite{cox2002transition}.

In addition to the Hele-Shaw arrangement, foams have also been studied in a number of other confining geometries. An example of this is the injection of monodisperse bubbles into a wedge-like geometry. In this case, the increasing plate separation forces the bubbles to adopt a initial monolayer structure which gives way to a bilayer (i.e. double-layer) and subsequently a multilayer arrangement \cite{Drenckhan:2005wk}. Variations in plate separation can also be used to control the rheology of highly ordered foams in microfluidic devices; strategically placed ``bumps'' in the channel force rows of bubbles to swap positions  \cite{Drenckhan:2005wk, Weaire:2008tk}. Other notable examples include foams in narrow cylindrical channels where the geometry compels the bubbles to spontaneously self organise into various helical arrangements \cite{boltenhagen1998giant, meagher2015experimental, pittet1996structural, Saadatfar:2008gv}. Bubble statistics and bubble dynamics of a foam trapped between narrowly separated concentric spheres have also been studied, as a direct means of testing the modified Von Neumann law for coarsening in the presence of curvature \cite{roth2012coarsening}.

\begin{figure} 
\begin{center}
\centering
\includegraphics[width=0.7\columnwidth ]{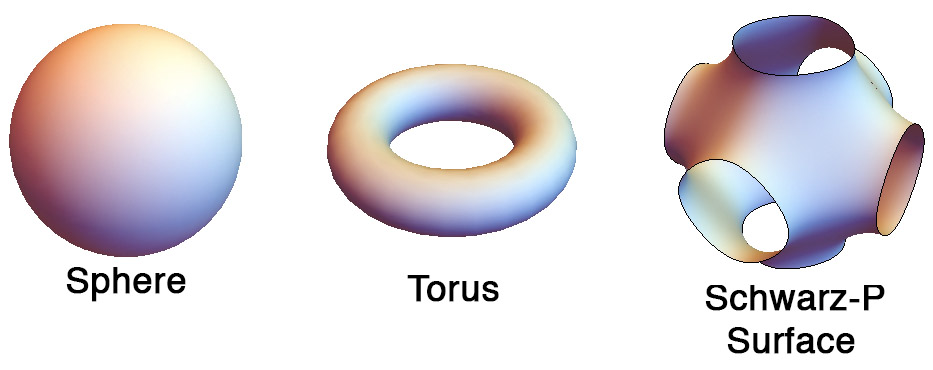}
\caption{The three different level-set surfaces considered in our numerical scheme.}
\label{three}
\end{center}
\end{figure}

Here - using the Surface Evolver package \cite{brakke1992surface}
- we present a framework for simulating individual bubbles, and multicellular dry foams, that are confined between narrowly separated curved plates. The bounding surfaces are modelled using level-set functions and the resulting bubble morphology is then governed by the minimisation of surface tension energy. Where possible we make use of content integrals (see below) to reduce the number of simulation elements required, thereby further reducing the numerical burden. We consider the following three cases:
\begin{itemize}
  \item  bubbles between concentric spherical plates (i.e. surfaces of constant positive Gaussian curvature);
 \item  bubbles between concentric tori (an arrangement with both positive and negative Gaussian curvature); and,
  \item  foams between confined between two Schwarz Primitive surfaces (an important example of a surface with negative Gaussian curvature),
\end{itemize}
the basic geometry of each is shown in Fig \ref{three}. In the first two cases the bounding surfaces are curved and parallel to each other. 
By parallel we mean that the two plates are parallel surfaces of each other. Thus if the points ${\bf p}$ on the first surface $I_1$ are translated along the surface normal ${\bf n}$ by a distance $d$ we obtain the second bounding plate \cite{do1976differential},
\[
I_2=\{{\bf p}+d\cdot {\bf n}({\bf p}) | {\bf p} \in I_1 \}.
\] 
However, in the third case (i.e. the P surface) it is not possible to define a pair of parallel surfaces in this manner. Below we describe the use of adjacent level set surface of the nodal approximation to the P surface, which we use to confine the bubble. However, this case will be more complex - compared to the spherical and toroidal cases - as variations in separation are anticipated to have significant consequences in determining the equilibrium morphology of a bubble (or foam) trapped between the substrates \cite{mughal2016}.

There are clear analogies between cellular structures on curved surfaces (whether they are foams or honeycombs) and previous experimental attempts to pack particles on curved surfaces \cite{bausch2003grain, altschuler2005global, irvine2010pleats}. In the latter case the curvature of the bounding surfaces induces a geometric frustration in the local crystalline order. This frustration is relieved by the presence of topological defects including disclinations, dislocations and more complex scar-like arrangements. We suppose that a similar mechanism may also govern the behaviour of quasi-2D foams on curved surfaces. Where, such defects have already been observed in the case of strictly 2D foams on a sphere \cite{cox2010minimal}. 

However, the case of quasi-2D foams between curved plates may prove to be richer than either that of particles or strictly 2D-foams on curved surfaces. Recently, we have shown that for a single bubble between a pair of parallel curved surfaces the surface tension energy of the bubble is sensitive to the Gaussian curvature of the bounding plates \cite{mughal2016}. A bubble has a low surface tension energy when it is in a region of positive Gaussian curvature and a high energy in a region of negative curvature. This energy difference can drive bubbles from a region of negative to positive curvature. Thus, at least in the case of soap froths there is an additional force or potential as compared to the problem of packing particles on a curved surface.

Here we demonstrate some of the technical detail how curved interfaces can be represented by level set constraints in the Surface Evolver. We discuss the instructive example of a single bubble on a sphere, followed by a discussion of the toroidal geometry as an example of a complex geometry with negative and positive curved regions. We then discuss the generation of multicellular foams on the Schwarz Primitive surface using nodal representations, and a Voronoi method to create the initial non-optimised partition.

\section{Foam Model}

A bubble consists of a liquid interface with a surface tension $\gamma$ enclosing a volume of gas $V$. The total surface energy of the bubble is given by,
\[
E=\gamma A
\]
where $\gamma$ is assumed to be constant and $A$ is the surface area of the bubble.

We assume that on two sides the bubble is bounded by solid surfaces, called substrates. The substrates are not necessarily flat but are smooth and frictionless. This allows the bubble interface to slide along the walls and relax to equilibrium. An interface makes a contact with angle $\theta_c $ with the wall given by Young's law:
\[
\gamma \cos \theta_c + \gamma_{S1} - \gamma_{S2} = 0,
\]
where $\gamma_1$ and $\gamma_2$ is the surface tension on the two sides of the interface. We assume the soap solution wets the solid surface and that the wetting films on both sides have same surface tension. Since both sides of the film contain the same gas, so that $\gamma_{S1} = \gamma_{S2}$, this gives $\cos \theta = 0$. As a result the soap film meets the confining wall at right angles (normal incidence). Furthermore, since $\gamma_{S1} = \gamma_{S2}$ the surface tension energy of the bubble depends entirely on the surface area of the (transverse) film between the walls, the wetting films in contact with the bounding surfaces make no contribution \cite{mancini2005equilibrium}.

\section{Surface Evolver simulations}

Simulations are conducted using the Surface Evolver package \cite{brakke1992surface}, which is an interactive \emph{finite element} program for the study of interfaces shaped by surface tension. Bubbles are represented by ``bodies'' which are comprised of oriented faces which are broken down into edges, which are in turn defined by vertices.  

Below we give details for modeling bubbles trapped between various plate geometries. The Evolver can handle such boundaries by constraining vertices to lie on level sets functions specified by the user in the input data file \cite{brakke1994surface}. The data files (below) consist of an initial starting geometry and level set constraints that model the walls. In some cases the geometry of the bubble is further simplified through the use of \emph{content integrals} (as described immediately below). Once the initial geometry of the bubble has been defined, the energy of the bubble is then minimised by applying gradient descent (and other methods such as conjugate gradient) while repeatedly refining the mesh to improve accuracy. 

The wetting surfaces of the bubble (i.e. the surfaces in contact with the bounding substrates) do not contribute to the surface tension energy of the bubble. As such the presence of these wetting surfaces in the Surface Evolver model represents an unnecessary numerical burden which can be removed. The problem then remains how to compute the volume of the the resulting open body (i.e. the numerical representation of the bubble)?  The solution is to compute the volume by a closed contour integral over the edges of the missing surface - such integrals are known as \emph{content integrals} and are described fully in the Surface Evolver manual \cite{brakke1994surface}. In two cases below (the spherical and toroidal geometries) we are able to take advantage of the cylindrical symmetry of the situation and derive the appropriate content integrals. As such the bubble model in these cases consists solely of the transverse film between the walls. For a simple example of the use of content integrals for a bubble trapped between two horizontal \emph{flat} plates see the canonical example \emph{plates}\_\emph{column.fe} \cite{wshop}.

\section{Bubbles between concentric spherical plates}

\begin{figure} 
\begin{center}
\centering
\includegraphics[width=0.3\columnwidth ]{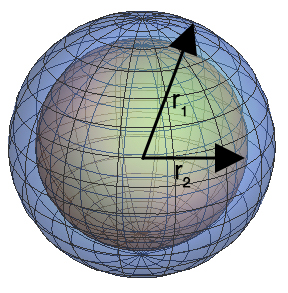}
\caption{Level set constraints to confine a bubble between two concentric spherical plates - an outer plate of radius $r_2$ and an inner plate of radius $r_1$.}
\label{sphere}
\end{center}
\end{figure}

In spherical polar coordinates  $(\theta, \phi)$ with $0\leq \theta \leq \pi, 0\leq \phi \leq 2\pi$ the parametrisation, 
\begin{eqnarray}
x&=&\sin(\theta)\cos(\phi), \nonumber\\
y&=&\sin(\theta)\sin(\phi), \nonumber\\
z&=&\cos(\theta), \nonumber
\end{eqnarray}
describes a sphere as the locus of points $(x,y,z)$ that satisfy the equation $\Phi(x,y,z)=r$. The outer bounding plate is described by the equation $\Phi(x,y,z)=r_1$ and the inner plate is given by $\Phi(x,y,z)=r_2$ where $r_1>r_2$, as shown in Fig \ref{sphere}. Note, for this geometry the bounding substrates are curved but parallel to each other everywhere.

\subsection{Content Integrals}

\begin{figure} 
\begin{center}
\centering
\includegraphics[width=0.6\columnwidth ]{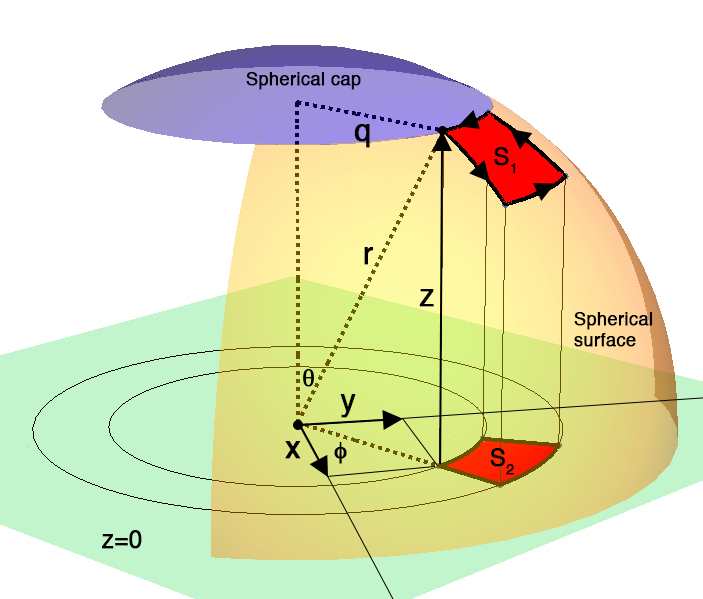}
\caption{Computing the area under a sphere of radius $r$ using the method of cylindrical shells}
\label{sphereint}
\end{center}
\end{figure}

Here we derive the content integral for a bubble between two concentric spheres. Consider the spherical cap of radius $q$ on a sphere of radius $r$ as shown in Fig \ref{sphereint}. We can integrate over a series of infinitesimally thin concentric cylinders, each of radius $q'$, length $z$ and area $A=(2\pi q') \;z$ to find the volume under the spherical cap, so that,
\begin{equation}
V=\int_{0}^{q} (2\pi q') \;z \; \textrm{d}q'= \int_{0}^{q} (2\pi q') \;\sqrt{r^2-q'^2} \; \textrm{d}q',
\label{eq:sv1}
\end{equation}
where we have used the relationship $r^2=z^2+q'^2$. Carrying out the integration gives,
\begin{equation}
V=\frac{2\pi}{3} \left( r^3 - (r^2-q^2)^\frac{3}{2} \right).
\label{eq:sv2}
\end{equation}
Since $q^2 = x^2 + y^2$ we have $r^2 - q^2 = z^2$ - and Eq. ($\!\!$~\ref{eq:sv2}) simplifies to,
\begin{equation}
V= \frac{2\pi}{3}(r^3 - z^3). 
\end{equation}
Here $V$ is the volume under the entire cap obtained by integrating over $\phi$, we can obtain the differential volume if we replace $2\pi$ by $\textrm{d}\phi$ to give,
\begin{equation}
\textrm{d}V = \frac{1}{3}(r^3 - z^3)d\phi.
\label{eq:dv1}
\end{equation}
Now, note that $\phi(x,y)=\tan^{-1}(y/x)$ so that,
\begin{eqnarray}
\textrm{d}\phi (x,y)
&=&
\frac{\partial \phi(x,y)}{\partial x} \textrm{d}x
+
\frac{\partial \phi(x,y)}{\partial y} \textrm{d}y,
\nonumber
\\
&=&
\frac{-ydx + xdy}{x^2 + y^2},
\nonumber
\\
&=&
\frac{-ydx + xdy}{r^2 - z^2},
\label{eq:dphi}
\end{eqnarray}
where we have used the identity $x^2+y^2+z^2=r^2$ in the last step. Upon substituting Eq. ($\!\!$~\ref{eq:dphi}) into Eq. ($\!\!$~\ref{eq:dv1}) we have,
\begin{equation}
\textrm{d}V 
= 
\frac{1}{3}(r^3 - z^3)\frac{(-ydx + xdy)}{(r^2-z^2)},
\end{equation}
or cancelling the common factor $(r-z)$, gives,
\begin{equation}
\textrm{d}V 
= 
\frac{1}{3} \frac{r^2 +rz + z^2}{r+z}(-ydx + xdy).
\end{equation}
The volume under the sphere between the patches labelled $S_1$ and $S_2$ is then a contour integral $V=\oint_C dV$ - that can be evaluated by tracing a positively orientated closed path along the edges of the patch $S_1$, as shown in Fig \ref{sphereint}. Note the above integrand has no singularity at the north pole $z=r$. It does have a singularity at the south pole, and so {\bf must not} be used near the south pole.

In the case of a bubble between two spherical plates the Evolver computes two such integrals one on the inner sphere and one on the outer sphere (see data file below). Since the surface normals on these two constraints point in the opposite directions the resulting volume is arrived at by the difference between these two volume integrals.

\medskip
\medskip

The use of the spherical constraints and the content integral is demonstrated in the following Surface Evolver data file.

\subsection{Sphere.fe}
\medskip

\begin{verbatim}
parameter r1=20.5 /*Radius of outer sphere*/
parameter r2=20.0 /*Radius of inner sphere*/
 
constraint 1 /* The outer spherical plate */ 
formula: x^2 + y^2 + z^2 = r1^2
content:  /* Sphere volume element */
c1: -(r1^2 +r1*z + z^2)*(-y)/(r1+z)/3
c2: -(r1^2 +r1*z + z^2)*(x)/(r1+z)/3
c3: 0

constraint 2 /* The inner spherical plate */ 
formula: x^2 + y^2 + z^2 = r2^2
content: /* Sphere volume element */
c1: -(r2^2 +r2*z + z^2)*(-y)/(r2+z)/3
c2: -(r2^2 +r2*z + z^2)*(x)/(r2+z)/3
c3: 0

function real zz1 ( real xx, real yy )
   { 
        return sqrt( (r1^2) - ((xx*xx) + (yy*yy)) )
   }

function real zz2 ( real xx, real yy )
   { 
        return sqrt( (r2^2)- ((xx*xx) + (yy*yy)) )
   }

vertices
1       1.0 1.0 zz2(1.0, 1.0) constraint 2
2       2.0 1.0 zz2(2.0, 1.0) constraint 2
3       2.0 2.0 zz2(2.0, 2.0) constraint 2 
4       1.0 2.0 zz2(1.0, 2.0) constraint 2

5       1.0 1.0  zz1(1.0, 1.0) constraint 1
6       2.0 1.0  zz1(2.0, 1.0) constraint 1
7       2.0 2.0  zz1(2.0, 2.0) constraint 1
8       1.0 2.0  zz1(1.0, 2.0) constraint 1

edges
1       1 2 constraint 2 
2       2 3 constraint 2 
3       3 4 constraint 2 
4       4 1 constraint 2 

5       1 5 
6       2 6 
7       3 7 
8       4 8 

9       5 6 constraint 1 
10      6 7 constraint 1 
11      7 8 constraint 1 
12      8 5 constraint 1 

faces
1       1 6 -9 -5 
2       2 7 -10 -6  
3       3 8 -11 -7  
4       4 5 -12 -8

bodies 
1    1 2 3 4 volume 0.5

\end{verbatim}

\section{Bubbles between concentric toroidal plates}

\begin{figure} 
\begin{center}
\centering
\includegraphics[width=1.0\columnwidth ]{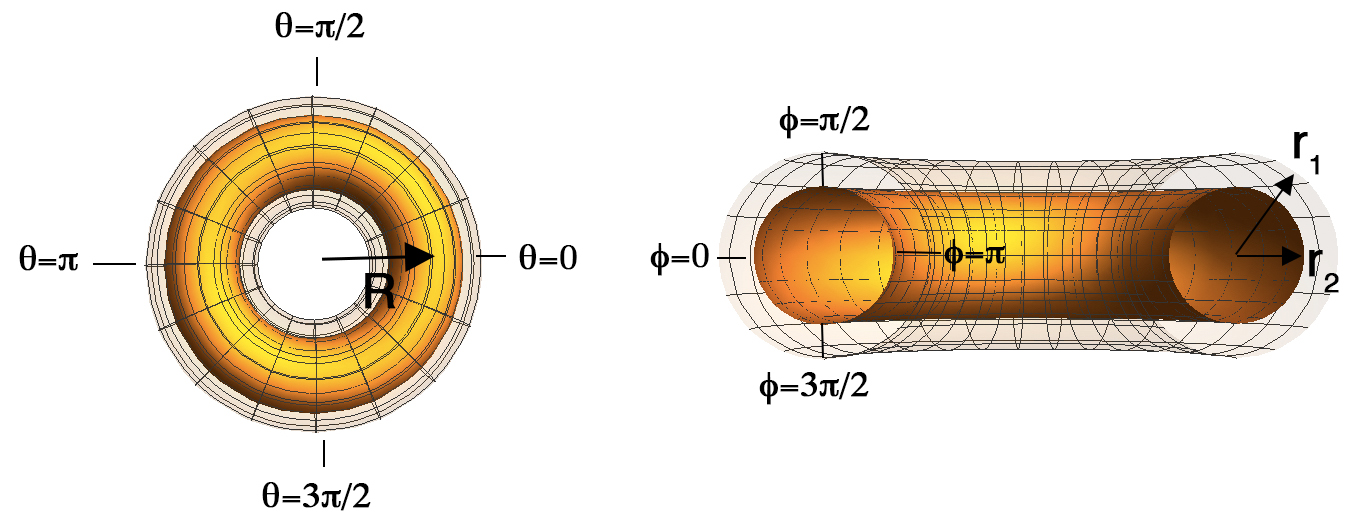}
\caption{Level set constraints to confine a bubble between two concentric toroidal plates - both tori have radius $R$. The outer torus has a tube radius $r_2$ while the inner torus has a tube radius $r_1$.}
\label{torus}
\end{center}
\end{figure}

In the angular coordinates $(\theta, \phi)$ with $0\leq \theta \leq 2\pi, 0\leq \phi \leq 2\pi$ the parametrisation,
\begin{eqnarray}
x&=&(R+r\cos(\phi))\cos(\theta), \nonumber\\
y&=&(R+r\cos(\phi))\sin(\theta), \nonumber\\
z&=&r\sin(\phi), \nonumber
\end{eqnarray}
defines a torus as the locus of points $(x,y,z)$ that satisfy the equation $\Phi(x,y,z,R)=r$
where,
\[
\Phi(x,y,z,R)=\sqrt{( R-\sqrt{x^2 + y^2 } )^2 + z^2}.
\]
The centre line of the torus is a circle of radius $R$ centred on the origin and the toroidal tube itself has a radius $r$ - see Fig \ref{torus}. This defines the inner toroidal plate while the outer plate is described by $\Phi(x,y,z,R)=r+d$, where $d$ is again the gap width. Note, that as in the spherical case, the bounding substrates are curved but remain strictly parallel to each other everywhere.

\subsection{Content Integrals}

\begin{figure} 
\begin{center}
\centering
\includegraphics[width=0.7\columnwidth ]{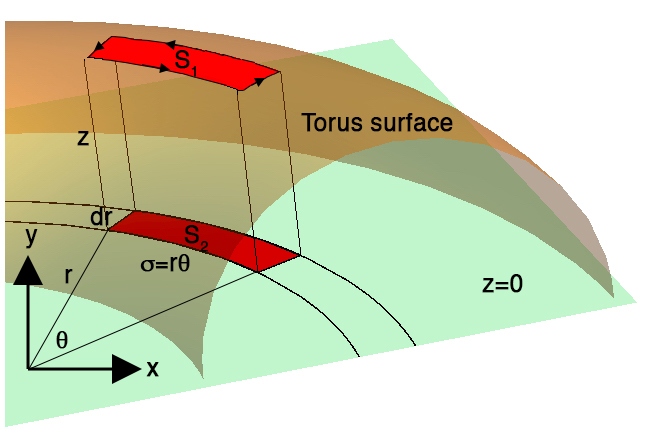}
\caption{Computing the area under a torus using the method of cylindrical shells.}
\label{torusint}
\end{center}
\end{figure}

Here we derive the content integral for a bubble trapped between two toroidal substrates. Consider for example the patch $S_1$ on the surface of a torus, as shown Fig \ref{torusint}, and another patch $S_2$ on the plane $z=0$, where $S_2$ is generated by projecting $S_1$ on to the plane. Here we shall calculate the volume of the box between the two patches using the method of cylindrical shells.

As shown in Fig \ref{torusint} we can let the radial distance along the plane $z=0$ be given by $r = \sqrt{x^2+y^2}$. Then the resulting arc of radius $r$ and angular extent $\theta(x,y)=\tan^{-1}(y/x)$ has a length,
\begin{equation}
\sigma=r\theta=r\tan^{-1}(y/x).
\label{eq:arclen}
\end{equation}
Hence we can integrate over a series of cylindrical sections each of area $A=z \sigma$, thickness $d\textrm{r}$, and volume,
\begin{equation}
dV=z \sigma dr.
\label{eq:dvtorus}
\end{equation}
Where,
\begin{eqnarray}
\textrm{d} r(x,y)
&=&
\frac{\partial r (x,y)}{\partial x} \textrm{d}x +\frac{\partial r (x,y)}{\partial y} \textrm{d}y 
\nonumber
\\
&=&
\frac{xdx+ydy}{\sqrt{x^2+y^2}}
\nonumber
\\
&=&
\frac{xdx+ydy}{r}.
\label{eq:drelement}
\end{eqnarray}
Thus upon substitution of Eq. ($\!\!$~\ref{eq:arclen}) and Eq. ($\!\!$~\ref{eq:drelement}) into Eq. ($\!\!$~\ref{eq:dvtorus}) we have,
\begin{eqnarray}
dV
&=&
z
\left[
r\tan^{-1}(y/x)
\right]
\left[
\frac{xdx+ydy}{r}
\right],
\nonumber
\\
&=&
z\tan^{-1}\left(\frac{y}{x}\right)(xdx+ydy).
\end{eqnarray}
As in the previous (spherical) example: the volume under the torus between the patches labelled $S_1$ and $S_2$ can be evaluated by tracing a positively orientated path along the edges of the patch $S_1$, as shown in Fig \ref{torusint}. 

Again, in the data file (below) there are two integrals to be evaluated - one on the inner torus and one on the outer torus. The difference between these two gives the volume of the bubble.

\medskip
\medskip

The use of the toroidal constraints and the content integral is demonstrated in the following Surface Evolver data file.

\subsection{Torus.fe}
\medskip

\begin{verbatim}
parameter r1=1.0 /*Tube radius of outer torus*/
parameter r2=0.9 /*Tube radius of inner torus*/
parameter Ro=5.0  /*Toroidal radius*/
parameter ss=0.1

constraint 1   /* The outer toroidal plate */
formula: ( Ro-sqrt( x^2 + y^2 ) )^2 + z^2 - r1^2 = 0
content: /* Torus volume element */
c1: x*z*atan2(y,x)
c2: y*z*atan2(y,x)
c3: 0

constraint 2   /* The inner toroidal plate */
formula: ( Ro-sqrt( x^2 + y^2 ) )^2 + z^2 - r2^2 = 0
content: /* Torus volume element */
c1: x*z*atan2(y,x)
c2: y*z*atan2(y,x)
c3: 0

function real zz1 ( real xx, real yy )
   {
        return sqrt( (r1^2) - (Ro - sqrt((xx*xx) + (yy*yy)))^2 )
   }

function real zz2 ( real xx, real yy )
   {
        return sqrt( (r2^2) - (Ro - sqrt((xx*xx) + (yy*yy)))^2 )
   }

vertices
1       (Ro-ss)   0.0   zz2(Ro-ss, 0.0)   constraint 2
2       (Ro+ss)   0.0   zz2(Ro+ss, 0.0)   constraint 2
3       (Ro+ss)   ss    zz2(Ro+ss, 0.0)   constraint 2
4       (Ro-ss)   ss    zz2(Ro-ss, 0.0)   constraint 2

5       (Ro-ss)   0.0   zz1(Ro-ss, 0.0)   constraint 1
6       (Ro+ss)   0.0   zz1(Ro+ss, 0.0)   constraint 1
7       (Ro+ss)   ss   zz1(Ro+ss, 0.0)   constraint 1
8       (Ro-ss)   ss   zz1(Ro-ss, 0.0)   constraint 1


edges
1   1 2   constraint 2   
2   2 3   constraint 2   
3   3 4   constraint 2   
4   4 1   constraint 2   

9   1 5
10   2 6	
11   3 7	
12   4 8	

5   5 6   constraint 1
6   6 7   constraint 1
7   7 8   constraint 1
8   8 5   constraint 1

faces
1       1 10 -5 -9 
2       2 11 -6 -10  
3       3 12 -7 -11 
4       4 9 -8 -12

bodies 
1    1 2 3 4 volume 0.005

\end{verbatim}

\section{Bubbles between Schwarz-Primitive surfaces}

\begin{figure} 
\begin{center}
\centering
\includegraphics[width=0.7\columnwidth ]{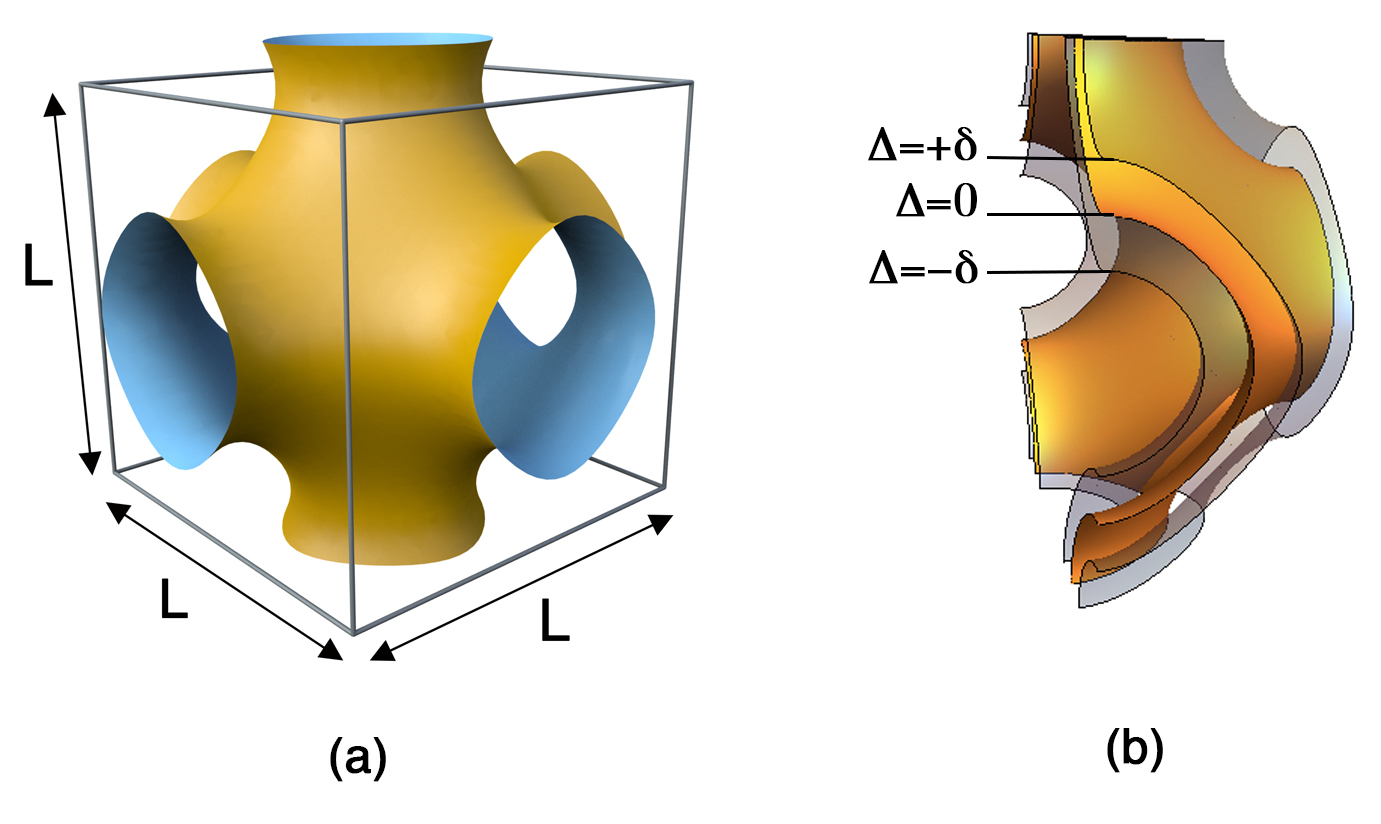}
\caption{(a) Nodal approximation of Schwarz P-Surface with periodicity L. (b) The middle surface $\Delta=0$ lies between the two substrates $\Delta=-\delta$ and $\Delta=+\delta$. For the purposes of the foam simulation (below) the $\Delta=0$ surface is a fictitious surface. The middle surface is used only by the simulating annealing routine to arrange the Voronoi centres evenly. In the foam simulation the surface $\Delta=0$ is discarded and only the substrates $\Delta=-\delta$ and $\Delta=+\delta$ are used.}
\label{single}
\end{center}
\end{figure}

Triply periodic minimal surfaces can be defined exactly using the Enneper-Weierstrass complex integration representation for some limited cases (including the Schwarz-P surface). While these methods have been used extensively to model triply periodic minimal surfaces, for our purposes they are unwieldy and instead we use the periodic nodal approximation of the Schwartz-P surface \cite{gandy2001nodal} ,
\[
\Phi=\cos(2\pi x/L) + \cos(2\pi y/L) + \cos(2\pi z/L)=\Delta,
\]
where $0 \leq x,y,z <L$, as shown in Fig \ref{single}. The surface is periodic in all three directions with a fundamental cubic unit cell of length $L$. In the following we shall consider the surface generated by setting the level-set $\Delta=0$ as well as adjacent surfaces  - for which $\Delta \neq 0$.

To generate a realistic foam structure on such a surface our method is as follows. N points, are distributed randomly over the P-Surface. These points represent initial coordinates for particles that move to minimise a repulsive inter-particle potential; in addition there is another potential which acts to keep the particles on the surface $\Delta=0$ (as described below). By this process the points are eventually evenly distributed over the surface $\Delta=0$. We then define two surfaces adjacent to $\Delta=0$, which are narrowly separated surfaces,
\[
\Phi_{\pm}=\cos(2\pi x/L) + \cos(2\pi y/L) + \cos(2\pi z/L)=\pm \delta,
\]
where $\delta << L$. A Voronoi partition of these N seed points is calculated. The resulting structure is the is then imported into the Surface Evolver, where each Voronoi cell represents a bubble. The bubbles are constrained by the bounding P-surfaces (using the constraints described below). The bubble areas are prescribed to be equal and we then use Evolver's gradient descent implementations to converge to a minimum area configuration.

\subsection{Simulated Annealing}
We begin first by distributing $N$ points randomly over the $\Delta=0$ surface (i.e. the approximation to the P surface), this initial arrangement is the starting point for Metropolis simulated annealing algorithm. 

The simulation is addressed to a three dimensional cube shaped cell of side length $L$. The box is periodic in all three direction. Contained within this space are $N$ points which represent the centres of $N$ softly repelling spheres, each of diameter $d$.  If a pair of spheres is sufficiently close that they overlap we account for this using a pairwise potential as described below. A second potential is used to force the sphere centres to lie on the P-surface. 

We model the overlap potential between spheres using a Hookean, or ``spring-like'', pairwise interaction between the $i$th and $j$th spheres, which have their centres at  ${\bf r}_i=(x_i, y_i, z_i)$ and ${\bf r}_j=(x_j, y_j, z_j)$, the interaction energy between spheres is then given by,
\begin{equation}
E^S_{ij}
 =
\left\{
\begin{array}{l l}
\frac{1}{2}(r_{ij}-d)^2 & \quad \mbox{if $r_{ij}\leq d$}\\
0 & \quad \mbox{if $r_{ij} >d$}\\
\end{array}
\right.
\end{equation}
where $r_{ij}=|{\bf r}_i-{\bf r}_j|$ is the distance between the centres of the spheres. Note the interaction energy  falls to zero when there is no overlap between the spheres.

For the $ith$ sphere a measure of its distance from the P Surface is given by
\[
\delta_i=\cos(2\pi x_i/L) + \cos(2\pi y_i/L) + \cos(2\pi z_i/L),
\]
If the sphere is on the implicit surface then $\delta_i=0$ and non-zero otherwise. From this we can associate an energy cost for the sphere if it is not on the surface as, 
\begin{equation}
E^P_{i}
=
\frac{1}{2}\delta_i^2.
\end{equation}
Thus the total energy of the system is given by the sum of the sphere-sphere and sphere-surface interactions. The energy of the system is then minimised by a Metropolis simulated annealing routine. This yields  a set of sphere centres that are evenly distributed over the surface $\Delta=0$ and with $E^P_{i}$ being negligibly small for any given centre.

\subsection{Voronoi Partition}

\begin{figure} 
\begin{center}
\centering
\includegraphics[width=1.0\columnwidth ]{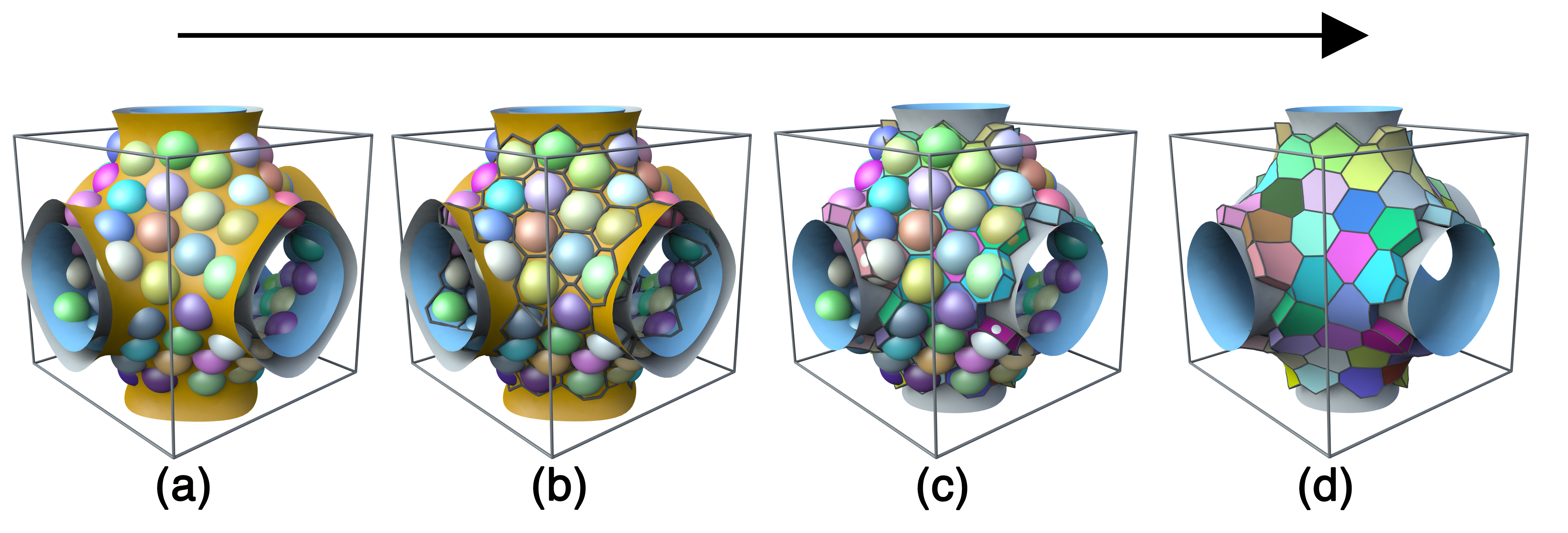}
\caption{Process to make the Voronoi partition. In the first step (a) a simulated annealing algorithm is used to evenly distribute sphere centres over the surface $\Delta=0$. Then (b) a Voronoi partition is generated from the sphere centres - the partition is clipped at the bounding surfaces and respects the local curvature of the substrates, as shown in (c). The final state (d) is then used as the starting point for a Surface Evolver simulation.}
\label{vprocess}
\end{center}
\end{figure}

As a result of simulated annealing all the particle centres lie close to the surface $\Delta=0$ and are distributed evenly over it. Before computing the Voronoi partition around the sphere centres, if any of the centres are not exactly on the surface $\Delta=0$ we project it onto the surface and use this set of adjusted points for the partition. The surface $\Delta=0$ is of no further interested and is deleted.

Next, we define two bounding surfaces $\Delta=+\delta$ and $\Delta=-\delta$, so that the spheres centres lie on the mid-surface between these boundaries, see Fig \ref{single}b and Fig \ref{vprocess}a. From these bounding surfaces and the sphere centres we compute the Voronoi partition which is periodic in all three directions, using the graphics package $Houdini$, see Fig \ref{vprocess}b. The resulting partition is clipped at the boundaries - i.e. the package calculates the confined Voronoi diagram that respects the local curvature of the bounding substrates, see Fig \ref{vprocess}c. The final state is shown in Fig \ref{vprocess}d.

The result is Voronoi cell around each particle represented in terms of the vertices of the Voronoi partition and POLYs. The latter are a data structure that consists of a closed positively orientated loop over the vertices. Each loop defines a face with a unit normal pointing to the exterior of the cell.

\subsection{Surface Evolver}

The Voronoi partition is then converted into the appropriate format for Surface Evolver simulations. Again, the bounding constraints are imposed through level set functions $\Delta=\pm \delta$. However, due to the lack of cylindrical symmetry the derivation of the content integrals is not attempted here. Instead we manually set the surface tension to zero on the edges and faces that are in contact with the bounding walls. An example of the data file for a simple bubble between two Schwarz-P surfaces is given below.

It should be noted (again) that unlike the previous examples, the separation between the adjacent level-set P surfaces is not a constant. Unlike the simpler cases (i.e. concentric spherical and toroidal plates) it is not possible to define a surface that is strictly parallel to the P surface everywhere. As such in addition to the effects of substrate curvature in determining the surface tension energy of the bubble there will be another analogous effect due to the effect of varying plate separation, see \cite{mughal2016} for further details.

\subsection{Psurface.fe}
\medskip
\begin{verbatim}
parameter delta=0.2

constraint 1   /* the outer plate */
formula: cos(2*pi*x)+cos(2*pi*y)+cos(2*pi*z) = delta

constraint 2  /* the inner plate */
formula: cos(2*pi*x)+cos(2*pi*y)+cos(2*pi*z) = -delta

function real zz1 ( real xx, real yy, real sl )
   { 
	return acos( sl  - cos(2*pi*xx) - cos(2*pi*yy) )/(2*pi);
   }

function real zz2 ( real xx, real yy, real sl )
   { 
    return acos( -sl  - cos(2*pi*xx) - cos(2*pi*yy) )/(2*pi);
   }

vertices
1   0.1  0.2  zz1(0.1, 0.2, delta)  constraint 1  /* vertices on outer plate */
2   0.2  0.1  zz1(0.2, 0.1, delta)  constraint 1
3   0.2  0.2  zz1(0.2, 0.2, delta)  constraint 1
4   0.15 0.25 zz2(0.15,0.25,delta)  constraint 2  /* vertices on inner plate */
5   0.25 0.15 zz2(0.25,0.15,delta)  constraint 2
6   0.25 0.25 zz2(0.25,0.25,delta)  constraint 2

edges  
1   1 2    constraint 1 
2   2 3    constraint 1 
3   3 1    constraint 1 

4   4 5    constraint 2 
5   5 6    constraint 2 
6   6 4    constraint 2 

7   1 4
8   2 5
9   3 6

faces  
1  1 2 3 constraint 1 tension 0 color yellow
2  4 5 6 constraint 2 tension 0 color yellow
3  7 -6 -9 3
4  -4 -7 1 8
5  -5 -8 2 9

bodies  
1   1 2 3 4 5   volume 0.01

read
re1 := {refine edges where on_constraint 1 }
re2 := {refine edges where on_constraint 2 }

// Typical (but crude) evolution
gogo := { re1; re2; g 5; r; g 20; r; g 50; r; g 50 } 

\end{verbatim}

\section{3-Bee Printing}

In the following we describe our experimental attempt to print a honeycomb onto a Schwarz-P surface using bees. 

\begin{figure} 
\begin{center}
\centering
\includegraphics[width=1.0\columnwidth ]{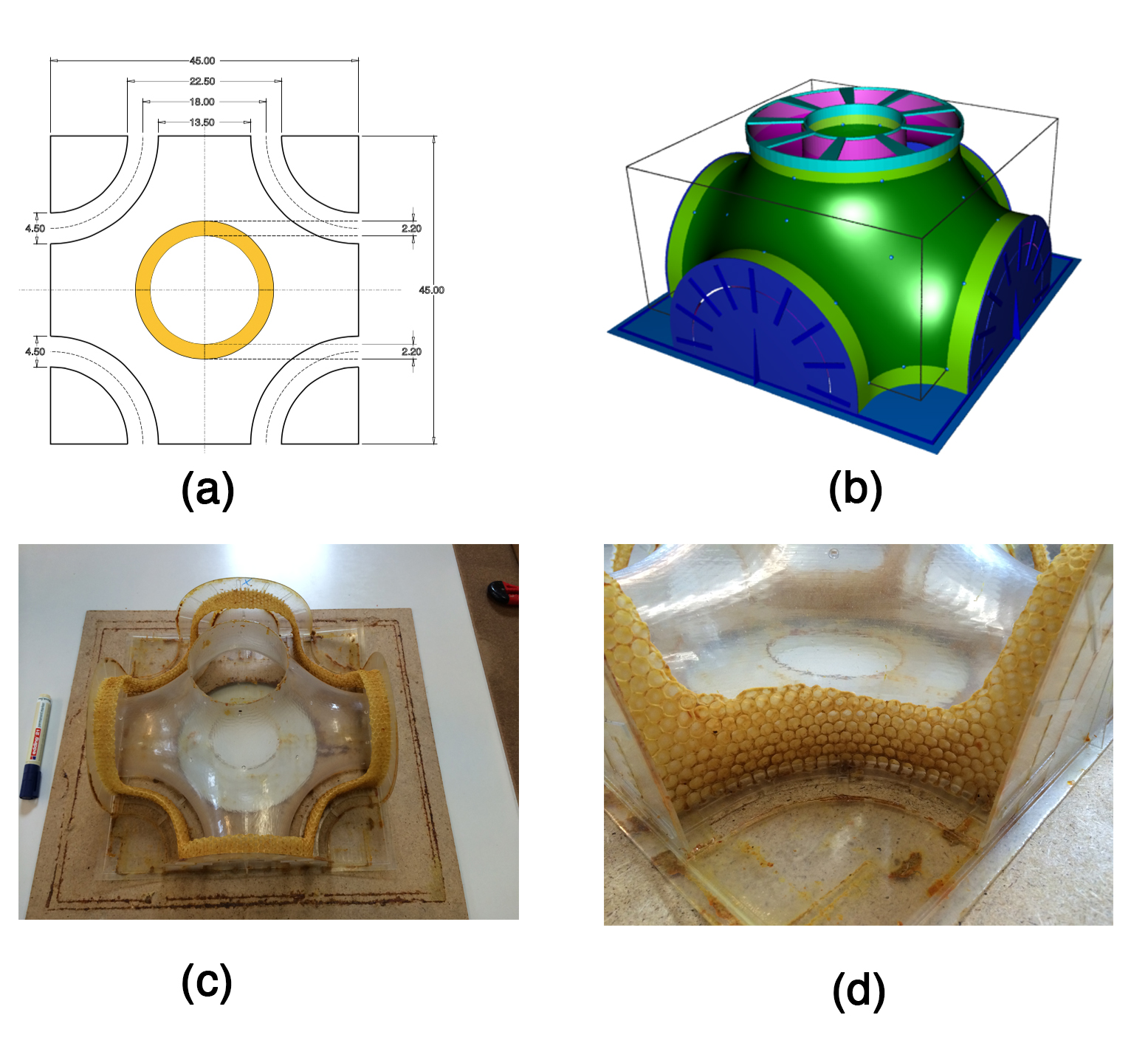}
\caption{(a) Schematic design of the cell - all distances are in centimeteres. (b) Complete cell for rapid prototyping. (c)  Result after a hive of bees is allowed to enter the structure and build on the surfaces. (d) Close up from previous image.}
\label{bees}
\end{center}
\end{figure}

Our motivation was to analyse how the bees cope with or respond to negatively curved surfaces that have two essential properties. Firstly, negatively curved surfaces (such as the Schwarz-P surface) can never have constant Gaussian and constant mean curvature. Therefore, the bees would experience frustration in building the honeycomb - since at different points of the surface they encounter different surface curvatures. Secondly, It is not possible to tile a negatively-curved surface with hexagons only and so we were curious to see how the bees would modify the usual hexagonal motif to adapt to this constraint. 

With this in mind we designed an enclosure comprised of two Schwarz-P surfaces with a small gap between them (of the type described above), see Fig \ref{bees}a. Our hope was that the bees would then build a natural two-sided honeycomb in-between these two surface - that is one layer of cells are formed on the inner surface and another layer on the outer surface. The gap width was carefully chosen to be that of the height of two honeycomb cells and a little extra space for the bees to manoeuvre within the gap between cells from adjacent surfaces. 

We deliberately used only half the Schwarz-P surface for the cell. With this design we did not need to use pins or other spacers between the surfaces to hold them at the appropriate distance. As such the bees had maximum freedom to build in any way they chose without obstructions due to the design of the cell.

The cell was then constructed by rapid prototyping Fig \ref{bees}b and the bees were allowed to enter the structure. The result after after repeated attempts was that the bees did build a honeycomb on part of the structure - see Fig \ref{bees}c and Fig \ref{bees}d. However, we did not obtain a complete tiling of the two surfaces. Ultimately, we believe that the strongly curved surface proved to much of a frustration to the bees and they only attempted to build on the regions that were immediately accessible and neglected to proceed further inwards. 

\section{Conclusion}
We have presented some of our on-going work to simulate foams between curved surfaces. In the future we will implement these numerical techniques and examine the role of curvature in determining the morphology and dynamics of quasi-2D foams. We also hope to repeat the bee experiment using the insight gained from our first attempt.

\section{Acknowledgements}
We are greatly indebted to Ken Brakke (the inventor of Surface Evolver) for all his kind and patient help in implementing the content integrals. AM acknowledges support from the Aberystwyth University Research Fund.

\newpage



%% References


\bibliographystyle{model1-num-names}






% \end{thebibliography}

\end{document}